\documentclass[fleqn,12pt,twoside]{article}
\usepackage{espcrc1}

\usepackage{graphicx}

\newcommand{\AmS}{{\protect\the\textfont2
  A\kern-.1667em\lower.5ex\hbox{M}\kern-.125emS}}

\title{Excited Baryons in Large $N_c$ QCD}

\author{Thomas D. Cohen \address[UMCP]{Department of Physics,
University of Maryland \\
        College Park, MD 20742, USA}%
        \thanks{Support from the US Department of Energy is gratefully acknowledged.}
       }

\begin{document}

\maketitle

\begin{abstract}
This talk reviews recent developments in the use of large $N_c$
QCD in the description of baryonic resonances.  The emphasis is on
the model-independent nature of the approach.  Key issues
discussed include the spin-flavor symmetry which emerges at large
$N_c$ and the direct use of scattering observables.  The
connection to quark model approaches is stressed.

\end{abstract}

\section{Introduction}\label{sec:intro}

This talk discusses some recent developments in the description of
excited baryons via large $N_c$
QCD\cite{CL,CLpenta,CDLN,CDLN2,CLsu3,phot}. The work discussed was
principally done in collaboration with Rich Lebed of ASU; two
students of mine, Abhi Nellore and Dan Dakin also made substantial
contributions and Dan Martin, a student of Rich Lebed's, also
contributed. This talk will be very light on the technical
details. I refer you to the original papers for a more detailed
description.

To begin with let me motivate why one should look at large $N_c$
QCD in this context.  The answer is quite simple. Consider the
three great lies:
\begin{itemize}
\item  The check is in the mail.
\item  Of course, darling, I will
respect you in the morning.
\item My model is based on QCD.
\end{itemize}
The perspective of this talk is that the third one of these lies
should be lumped with the first two.  Let us recall that the good
book, a.k.a the Particle Data Book \cite{PDG}, contains an immense
amount of information about baryons. But there is no simple
systematic way to compute the properties of these states directly
from QCD.  Virtually all workers in the field use models and the
constituent quark model is clearly the standard tool out there.
However, as will be discussed below, the constituent quark model
has many serious difficulties.  Thus it is very important to
develop model-independent ways to learn what we can about these
states even if these model-independent methods are highly limited.
Large $N_c$ QCD is such a model-independent tool.

\subsection{The Quark Model}

The word ``quark'' has three meanings.
\begin{itemize}
\item A nonsense word from James Joyce's {\it Finnegan's Wake} :
``Three quarks for Muster Mark.''
\item A fundamental degree of
freedom in QCD.
\item An effective degree of freedom in the quark
model (aliases: the constituent quark model, the naive quark
model...)
\end{itemize}
In many ways the last two are no more closely related than the first two.

The constituent quark model has many problems.  In the first place
as given it yields stable excited states.  To give resonance
widths, additional and totally {\it ad hoc} dynamics are needed to
describe coupling to mesons or quark-antiquark pairs.  Thus
structure and dynamics are not treated on the same footing.  This
is highly problematic for descriptions of excited baryons which,
after all, are resonances seen in scattering experiments.
Secondly, the connection to QCD is totally obscure.  The QCD
quarks are simply different beasts then constituent quarks.  A QCD
quark has a mass of $\approx$ 5 MeV; a quark model quark has a
mass $\approx$ 300 MeV.  Indeed, the role of the quark model in
the history of physics is quite ironic.  It played a truly
essential role in the development of QCD but once QCD was
discovered there was no know way to derive the quark model.

Despite these conceptual problems the naive quark model remains
the standard picture with which most hadronic physicists think
about states, particularly excited states.  The reasons for this
are clear: The quark model is easy to think about---it is
patterned after well-understood atomic physics.  Direct QCD
calculations for excited states using lattice techniques are
extremely difficult and it will be a long while before reliable
lattice studies are available.  Finally, the quark model
works...sort of.  Not all states are well described and some
predicted states have not been observed.

\subsection{Introduction to Large $N_c$ QCD}

The problem with QCD at low momentum QCD is the absence of a
natural expansion parameter.  In 1973, just a year after the
formalization of QCD, `t Hooft proposed that QCD could be
generalized from $SU(3)$ to $SU(N_c)$ and that $1/N_c$ can then
serve as an expansion parameter\cite{Hoo}.  He developed a clever
double line diagrammatic method following the color flow and
showed that a formal limit exists in which $N_c \rightarrow
\infty$, $g \rightarrow 0$ with $g^2 N_c$ held fixed.  In this
expansion planar diagrams of gluons dominate with each nonplanar
gluon costing a factor of $N_c^{-2}$ and each quark loop counting
a factor of $N_c^{-1}$.  These diagrammatic rules have important
implications for correlation functions for operators with the
quantum numbers of mesons or glueballs and from these, $N_c$
scaling rules can be deduced.  Witten generalized the approach to
include baryons by arguing that a mean-field picture becomes
increasing well justified as $N_c \rightarrow \infty$ \cite{Wit}.
Formally this analysis yields, among other rules, the following
$N_c$ scaling rules:
\begin{eqnarray}
m_{\rm meson} & \sim & N_c^0 \; \; \; M_{\rm baryon} \sim N_c^1 \nonumber \\
\Gamma_{n-\rm meson} &\sim& N_c^{1-n/2}  \; \; \; \;
{\rm thus} \; \; \; \Gamma_{3\rm -meson} \sim N_c^{-1/2}  \nonumber \\
g_{\rm meson-baryon} & \sim& N_c^{1/2} \;\; g_{2 \, \rm
meson-baryon} \sim N_c^0
\label{scaling}
\end{eqnarray}
where $\Gamma_{n-\rm meson}$ is an n-meson vertex and $g_{\rm
meson-baryon}$ is a coupling constant of a meson to a baryon.

There are important phenomenological consequences from these
generic scaling rules which give a cartoon-like description of the
real world:
\begin{itemize}
\item Baryons are heavy compared to mesons.

\item Mesons are weakly interacting among themselves.

\item Baryons are strongly coupled to mesons but baryon-meson
scattering is order unity.

\item OZI rule is qualitatively understood (it becomes exact as
$N_c \rightarrow \infty$).

\item Dominance of two-meson decays when possible,

\item Explains non-existence of $qq\overline{q}\overline{q}$
exotics.  (They cannot bind at large $N_c$).

\item  Domination of meson and glueball tree graphs in effective
theory; helps justify Regge picture.

\item Requires the existence of hybrid mesons ({\it e.g.,} states
with quantum numbers of a quark-antiquark and valence glue).
\cite{hyb}
\end{itemize}

\subsection{Spin-Flavor Symmetry For Large $N_c$ Baryons}

The large $N_c$ analysis discussed so far has been generic in the
sense that the specific spin and flavor quantum numbers played no
special role.  As will be seen, for the case of baryons spin and
flavor play an exceptionally important role.  In particular, a
contracted $SU(2N_f)$  (where $N_f$ is the number of degenerate
light flavors) emerges for baryons at large $N_c$.  This symmetry
is closely related to the $SU(2N_f)$ symmetry of the simplest
version of the quark model.

The key idea in the derivation of the symmetry is large $N_c$
consistency\cite{GS,DM}.  Suppose one were studying pion-nucleon
scattering.  The contribution from the direct and crossed Born
graphs are proportional to $g_{pi N N}^2 \sim N_c$.  All non-Born
graphs (including the sum of iterated pion exchange) according to
Witten's counting rules scales as $N_c^0$, apparently yielding a
scattering amplitude which scales as $N_c^1$.  However, this
violates unitarity.  The only way to make the counting consistent
is if the Born terms are canceled by other baryons ({\it e.g.,}
the $\Delta$) which are degenerate with the nucleon (for $N_c
\rightarrow \infty$) and for which there is a conspiracy in the
coupling constants.

Such a  structure is possible only if the vertices satisfy a set
of commutation relations. It turns out that  this set of
commutation relations is the Lie algebra of contracted $SU(2N_f)$.
Thus, this algebra, and its associated group, becomes exact in the
large $N_c$ limit. In this write-up, I will spare you details and
refer you to the original papers.  The key things I would like to
stress, however, is that this group structure captures
considerable dynamical information.  The low-lying baryon states
must fall into nearly degenerate multiplets corresponding to
representations of the group.  Moreover, within the subspace of
the states in a representation there is a Wigner-Eckert theorem at
work so that {\it all} operators can be expressed as c-numbers
({\it i.e.} reduced matrix elements) times matrix elements of
generators which can be determined entirely from group theoretic
considerations. Corrections to this must be of relative order
$1/N_c$ or smaller.

Now as it happens all the representations of this contracted group
are infinite dimensional.  The lowest-lying members of the
representation are physical and the higher lying states are seen
as large $N_c$ artifacts.  The usual nucleon and $\Delta$ are
assumed to be in the lowest representation of the group---this is
the {\it only} model dependence of the approach.  Now it turns out
this representation ``looks like'' states in a naive quark model
(for large $N_c$) with all quarks in identical s-wave
orbitals\cite{DJM}.  They look like the quark model states in the
sense that there is a one-to-one mapping between the states in the
representation and states in the quark model with the same quantum
numbers.

This lowest-lying representation is composed of states which have
$I=J$ (where $I$ is the isospin).  The lowest two states are
identified as the nucleon and $\Delta$.  Higher-lying states are
seen as $1/N_c$ artifacts; the nucleon-$\Delta$ mass splitting
goes like $N_c^{-1}$.  The ratio of matrix elements of operators
between different states in the representation are fixed by
Clebsch-Gordan coefficients for the group up to corrections which
vanish at large $N_c$.  For example, at large $N_c$, $g_{\pi N
\Delta} = \frac{3}{2} g_{\pi N N}$.  In nature, this relation
holds to a few percent so the approach has real predictive power.
(In fact this particular comparison works as well as it does in
part because this particular relation is ``gold-plated'' in that
it does not acquire any correction correction at order
$1/N_c$\cite{DM,DJM}.

There is a technical trick introduced by Dashen, Jenkins and
Manohar to do the group theory simply\cite{DJM}.  One can map the
generators onto a quark model with all quarks in a single s-wave
orbital with the generators given by:
\begin{equation}
X_{i a}=\frac{1}{N_c} q^{+} \sigma_i \tau_a q \; \; \;\; \; \;    J_{i }= q^{+} \sigma_i q \; \; \;\; \; \;    T_{a }= q^{+} \tau_a q \; ;
\end{equation}
in the limit $N_c \rightarrow \infty$ these generators reproduce
the commutation relations of the algebra.  Thus, to calculate all
of the relevant Clebsches it is sufficient to compute the matrix
elements in the quark model.  It should be noted here that this
does {\it not} mean the dynamics is that quark model---this is
merely a ``poor man's'' way to do group theory.

One key result of the group theory which will play a crucial role
for the excited states is the $I=J$ rule.  This rule states that
all operators which contribute in leading order carry quantum
numbers with $I=J$.  This rule was originally seen in the Skyrme
model\cite{MM} but was subsequently derived directly from the
group theory of large $N_c$ QCD\cite{KM}.  Operators violating
this rule are suppressed by a factor of $N_c^{-|I-J|}$.

\section{Large $N_c$ and Excited Baryons}

At first sight it may seem straightforward to simply extend this
analysis directly to excited baryons.  If the states are fixed by
symmetry and the symmetry can be simply encoded in a quark model
basis it seems that one can simply create a large $N_c$ quark
model and automatically get the large $N_c$ results.  In fact, a
number of papers have done just
that\cite{PY,CGO,Leb,Leb2,CGKM,G,SGS,CC,CC2}.  There  are two
conceptual problems associated with this, however.  The first is
that the derivation of the group theoretic result was only for the
space states degenerate with the nucleon at large $N_c$ and these
excited states are split from the nucleon by order $N_c^0$.  Thus,
{\it a priori} there is no direct justification for the group
structure without redoing the large $N_c$ consistency argument.
Now as it happens Pirjol and Yan in a technical {\it tour de
force} did precisely this and showed that a contracted $SU(2N_f)$
structure arises for the excited states \cite{PY}.

Despite the beautiful mathematics of ref.~\cite{PY} it has an
underlying conceptual problem: it is based on the scattering of
pions off of asymptotic baryon states.  However, the excited
baryons are not generically stable asymptotic states even at large
$N_c$.  As shown originally by Witten the characteristic width of
an excited baryon is $N_c^0$---they do not become stable at large
$N_c$\cite{Wit}.  Thus, it is not legitimate to do an analysis
based on the assumption that the scattering amplitudes of a pion
off of an excited baryon is well defined.  This problem could be
evaded at least for some class of states if there existed a class
of states which for symmetry reasons has a width which goes like
$N_c^{-1}$.  Now as it happens, in reference \cite{PY} it is shown
that in the context of a simple large $N_c$ quark model there are
states with a width $\sim N_c^{-1}$.  However, as discussed in
ref.~\cite{CDLN}, this is only an artifact of the simple quark
model and is not a generic large $N_c$ result.  Thus the
straightforward extension of the techniques for the ground band
are not justified for excited states.

\subsection{Scattering Amplitudes}

The key idea that Rich Lebed and I pursued to evade the difficulty
that a straightforward extension of the techniques used for the
ground band to excited states was to focus directly on physical
observables rather than on particular baryon ``states''.  Recall,
that difficulty with focusing on particular excited baryon states
is that they are resonances rather than stable states.  Moreover,
as a practical experimental matter, the only way we know about
these state is through scattering experiments.  Thus, our first
goal is to use large $N_c$ methods to understand scattering.  As a
matter of principle we do not know directly from large $N_c$
methods whether QCD has any baryon resonances (or more to the
point, any baryon resonances which are narrow enough to observe
even if we lived in a truly large $N_c$ world).  However, if we
are able to relate scattering amplitudes in different channels to
each other at large $N_c$, then we can conclude that if there is a
pole in the scattering amplitude at complex energy---{\it i.e.,} a
resonance---in one channel it will have a nearby partner in
another.

To see how this works consider for simplicity the case of
two-flavored QCD and focus on pion-baryon scattering where the
baryon is a ground state baryon (nucleon, $\Delta$).  A generic
scattering amplitude can be labeled $S_{LL^\prime R R^\prime
IJ}^\pi$, where $L$ ($L'$) is the initial (final) pion orbital
angular momentum, $R$ ($R'$) is the initial (final) spin and
angular momentum of the baryon, and $I_s$, $J_s$ is the total
isospin and angular momentum in the s-channel.  The key thing is
that this amplitude is an operator in the space of baryons and,
hence, at leading order satisfies the $I=J$ rule (which as a
scattering process is a t-channel variable).  Since general
isospin and angular momentum allows more t-channel amplitudes than
the $I=J$ rule does, at large $N_c$ QCD the various amplitudes are
related.  Using standard recoupling identities to go from the
t-channel to s-channel \cite{CL} yields the following large $N_c$
result:
\begin{eqnarray}
S_{LL^\prime R R^\prime IJ}^\pi & = &\sum_K (-1)^{R^\prime - R}
\sqrt{(2R+1)(2R^\prime+1)} \nonumber\\
&{}&(2K+1) \left\{ \begin{array}{ccc} K & I & J\\
R^\prime & L^\prime & 1 \end{array} \right\} \left\{
\begin{array}{ccc} K & I & J \\ R & L & 1 \end{array} \right\}
s_{KL^\prime L}^\pi ,
\label{MPeqn1}
\end{eqnarray}
An analogous result can be derived for $\eta$ nucleon scattering.
\begin{equation} S_{L R J}^\eta  =  \sum_K
\sqrt{(2R+1)(2R^\prime+1)} (2K+1)
\delta_{KL} \, \delta (LRJ) \, s_{K}^\eta .
\label{MPeqn2}
\end{equation}
The reason that various scattering amplitudes are linearly related
is clear from the structure of eqs.~(\ref{MPeqn1}) and
(\ref{MPeqn2}): There are more amplitudes $S_{LL^\prime R R^\prime
IJ}^\pi$ than there are $s_{K L^\prime L}^\pi$ amplitudes, and
thus there are linear constraints between them that hold to
leading order in the $1/N_c$ expansion; similarly, there are more
$S_{L R J}^\eta$ amplitudes than $s_{K}^\eta$ amplitudes.

Note that in both the $\pi$  and $\eta$ scattering cases there are
``reduced'' scattering amplitudes $s_{KL^\prime L}^\pi$ and
$s_{K}^\eta$.  Both of these depend on a variable $K$ which is
summed over in the final expression. As will be seen, this $K$
quantum can be used to distinguish resonances.

\begin{figure}[htb]

\includegraphics{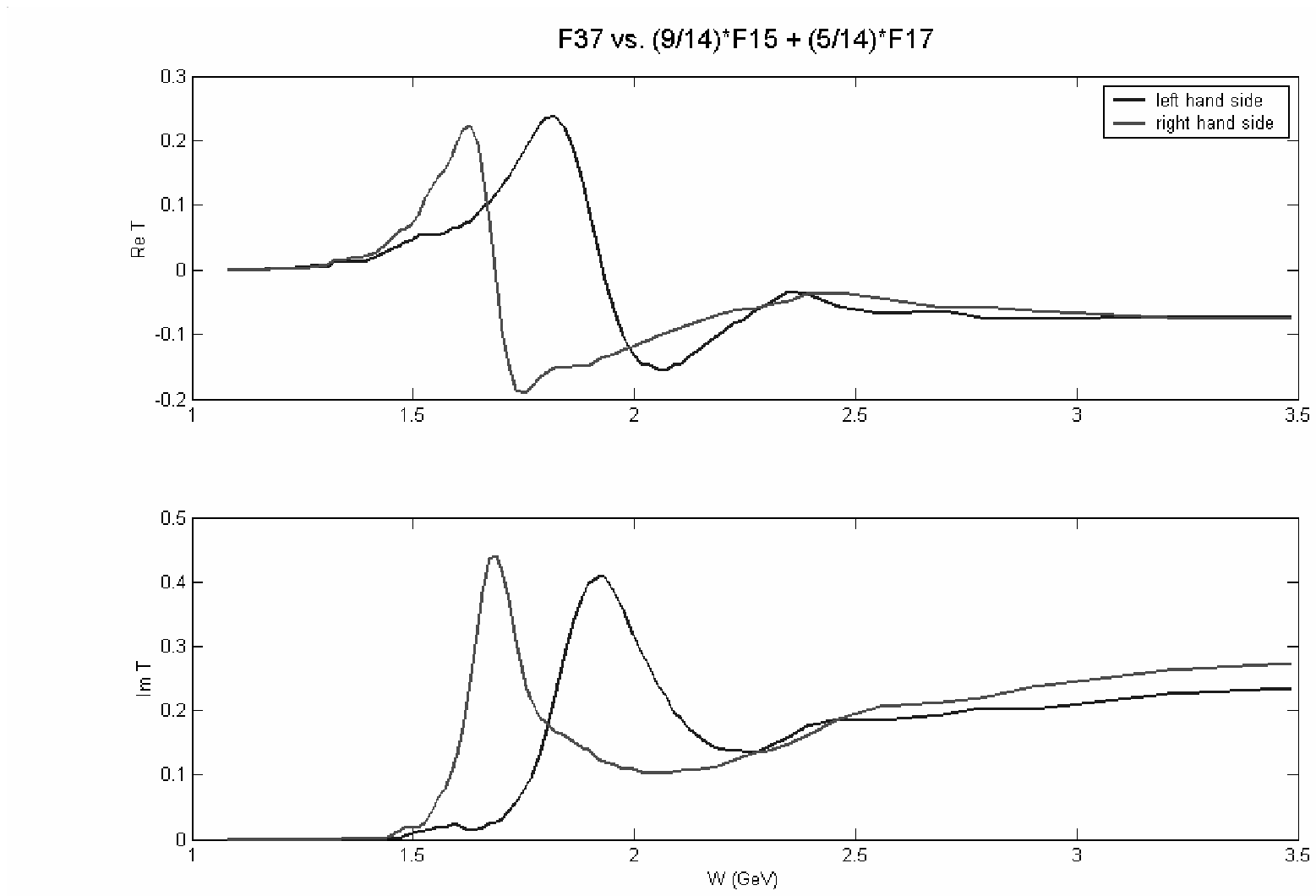}
\caption{Comparisons of the scattering amplitudes of the
right-hand side of eq.~(\ref{f}) with the left-hand side.  The
upper curve represents the real part while the lower curve
represents the imaginary.  The amplitudes are extracted from a
partial wave extraction of the fit to scattering data \cite{S}. }
\label{fig1}

\end{figure}

While these formulae are not new---they were previously derived in
the context of the Skyrme model\cite{Sk}---the present
model-independent derivation is. One can algebraically eliminate
the reduced matrix elements to obtain relations directly between
the physical amplitudes which hold to leading order in $1/N_c$. I
will refer you to the original papers for an enumeration of all
such relations and a test of how well they work.  Here for
concreteness I focus on one particularly illuminating case of
$\pi$-N scattering      ($R=R'=1/2$) with $L=L'=5$ (f-waves). The
scattering amplitudes can be denoted in the form $S_{F, 2(I+1),
2(J+1)}$
\begin{equation}
S_{f,3,7}=\frac{9}{14} S_{f,1,5} + \frac{5)}{14} S_{f,1,7}
\;\label{f}
\end{equation}
In fig.~\ref{fig1} the left and right
sides of eq. (\ref{f}) are plotted for both the real and the
imaginary parts of amplitudes extracted from the scattering data.
It is easy to envision how the two sides can overlap exactly as
$N_c \rightarrow \infty$.

Apart from the fact that they demonstrate the predictive power of
the method, the data in fig. \ref{fig1} also show how resonances
are related.  Clearly there is a resonance in the F37 channel,
since the left-hand side must equal the right-hand side up to
$1/N_c$ corrections there clearly must also be a resonance in at
least one of the channels on the right-hand side and that is
exactly what we see.  Thus, although I have not shown from first
principles that either channel needs to have a resonance, I have
demonstrated that {\it if} there is a resonance in one channel
there will be resonances in other channels which are degenerate up
to $1/N_c$ corrections.

Before completing a discussion of the degeneracy of resonances and
the symmetries that it entails, it is useful to mention at this
stage that the method can be generalized and extended.  For
example one can work at next to leading order (which only has
additional  predictive power in the reaction $\pi {\rm N}
\rightarrow \pi \Delta$)\cite{CDLN} or study $\eta$N scattering,
photoprodction reactions\cite{phot} or scattering in exotic
(pentaquark) channels\cite{CLpenta}.

\subsection{Baryon Resonances}

A resonance is a pole in the scattering amplitude at unphysical
kinematics.  Suppose for the sake of argument that
eq.~(\ref{MPeqn1}) was exact ($1/N_c$ corrections were negligible)
and that there was a resonance in a particular channel and, hence,
a pole in the scattering amplitude at some unphysical point.  Such
a divergence is only possible if one of the reduced amplitudes
itself diverges.  Therefore, the existence of a resonance implies
a pole in a reduced amplitude.  However, a single reduced
amplitude contributes to many physical amplitudes.  Thus the
existence of a resonance in one channel in large $N_c$ QCD
predicts the existence of a resonance in other channels. Note that
not only does this imply the existence of a resonance in the other
channels but the existence of a resonance at the same unphysical
value for the energy.  This means that the resonances are
degenerate: large $N_c$ QCD predicts the existence of degenerate
multiplets of resonances.

A few comments are in order about such multiplets.  The first is
that position of the resonances for these multiplets are
degenerate in the complex plane: thus they are degenerate in both
mass and width.  This result needs to be taken with a grain of
salt, however, due to $1/N_c$ corrections.  While it is formally
true at large $N_c$ that the widths and masses will be degenerate,
one might expect that for the lowest-lying resonances of physical
interest that there might be large $1/N_c$ corrections to the
widths.  The reason for this is simply that  for low-lying states
the widths  are highly sensitive to the available phase space, and
small variations in the mass of state due to $1/N_c$ corrections
can yield large differences in the widths.  It is probably better
to characterize the widths and coupling constants as being nearly
degenerate at large but finite $N_c$ rather than the masses and
the widths.

Note that these multiplets share a common $L$, $L'$ and $K$. The
$L$ and $L'$ are important in characterizing the incoming and
outgoing channels while the $K$ quantum number really
characterizes the intrinsic properties of the resonances.  Note
the existence of degenerate multiplets of states is exactly what
one expects if the states fall into representations of some group.
In this sense the analysis above implies the existence of an
emergent $SU(2N_f)$ symmetry for the excited states at large $N_c$
as well as the ground states.  This provides some {\it a
posteriori} justification for the treatments of
refs.~\cite{PY,CGO,Leb,Leb2,CGKM,G,SGS,CC,CC2}.

It is important to understand the connection between this approach
and the quark model based approach of refs.
\cite{CGO,Leb,Leb2,CGKM,G,SGS,CC,CC2}.  On the one hand, as noted
above, there is no real {\it a priori} justification for the
approach.  Unlike the analysis of the ground band baryons, for
excited baryons the quark model approach is more than just ``poor
man's group theory''; it makes real dynamical assumptions
including the assumption that the states are stable (or at least
narrow at large $N_c$ and that some fixed number of quarks are in
excited orbitals).  Despite this,  there is a strong connection
between this approach and the model-independent method described
above. In particular, the underlying group structure of the quark
model implies that if one only includes the leading order
operators in the $1/N_c$ expansion then the  excitation energies
in the quark model fall neatly into degenerate
multiplets\cite{CGO,Leb,Leb2,CGKM,G,SGS,CC,CC2}.  This fact is not
manifestly clear in the treatments of
\cite{Leb,Leb2,CGKM,G,SGS,CC,CC2} since subleading operators were
included at the outset.  Moreover, the spin-flavor quantum numbers
of these multiplets are in a one-to-one correspondence with those
allowed for a particular $K$ multiplets resonant analysis
above\cite{CL}.  Thus, the quark model captures the multiplet
structure of the underlying $1/N_c$ dynamics.

\subsection{Phenomenological Consequence}

In the best of all possible worlds, one would see clear evidence
for the large $N_c$ multiplet structure in the hadronic data with
small splittings due to $1/N_c$ effects.  Alas, in this world
things are not so nice.  The difficulty is that splittings between
multiplets are small compared to the splitting within each
multiplet.  The reason for this is unclear but is not connected
with $1/N_c$ physics in any obvious way.  To make sense of this
situation more analysis is required.  A better understanding of
how higher-order effects enter is clearly needed and work along
this direction is being pursued.  Another possible way to see
effects is to broaden the analysis to three flavors and see if the
effects of the multiplet structure are more apparent there. Again,
to do this, systematic higher-order corrections are necessary
(this time in $SU(3)$ flavor breaking).  Analysis of the $SU(3)$
generalization has begun\cite{CLsu3}.

There is, however, one place where the large $N_c$ analysis has
already borne phenomenological fruit and that is the study of {\it
decays} of negative parity nucleon states. One puzzling aspect of
these states is the fact that the N(1520) decays very strongly to
the $\eta$N channel and comparatively weakly to the $\pi$N channel
(they have very similar branching fractions even though the phase
space for $\pi N$ is a factor of ~3 greater) while the N(1650)
decays very strongly to the $\pi$N channel and very weakly to the
$\eta$N channel. This can be easily understood from large $N_c$.
If one assumes that the states to good approximation fall into $K$
multiplets with small admixtures due to $1/N_c$ effects, then this
behavior is quite natural.  It is easy to see by tracking through
the quantum numbers that a pure $K=1$ negative parity nucleon
cannot decay into $\eta N$ and a pure $K=0$ negative parity
nucleon cannot decay into $\pi N$. Thus if one identifies the
N(1520) as (predominantly) a $K=0$ state and the N(1650) as
(predominantly) a $K=0$ state, the decay patterns are easily
understood.

\end{document}